\documentclass[12pt]{article}
\usepackage{a4}
\usepackage{psfig}
\usepackage{epsfig}
\usepackage{epsf}
\usepackage{rotating}
\usepackage{citesort}       
\usepackage{amssymb}
\usepackage{graphicx}

\newcommand{\lsim}{\mbox{$\leq$}}

\newcommand{\anueR}{\mbox{$\overline{\nu}_{eR}$}}

\newcommand{\text}[1]{\mbox{$\rm #1 $}}
\newcommand{\anureaction}{\mbox{$\overline{\nu}_e+p\to n+e^{+}$}}

\def\plb#1#2#3{    { Phys. Lett. }{\bf B #1} (19#2) #3}

\newcommand{\be}{\begin{equation}}
\newcommand{\ee}{\end{equation}}

\begin{document}
\phantom{peppe}
\vspace{-2cm}
\begin{flushright}
{\small
IFUM-346-FT, 
FT-UM-TH-03-15, 
IFTM-03-14
}
\end{flushright}

\vspace{1cm}

\begin{center}
{ \Large \bf  KamLAND,
neutrino transition magnetic moments
and the solar magnetic field}\\[0.2cm]
\large 
V.~Antonelli$^{a\star}$, 
Bhag C. Chauhan$^{b\star}$\footnote{On leave from Govt. Degree College, Karsog (H P) India 171304.}, 
Jo\~{a}o Pulido$^{b\star}$,
E.~Torrente-Lujan$^{c\star}$\\[2mm]
$^a$ {\small\sl Dip. di Fisica, Univ. di Milano,
 and INFN Sez. Milano,  Via Celoria 16, Milano, Italy\\
$^b$ 
Centro de F\'\i sica das Interac\c c\~oes Fundamentais (CFIF),
 Departamento de F\'\i sica, Instituto Superior T\'ecnico,
Av. Rovisco Pais, P-1049-001 Lisboa, Portugal\\
$^c$ Departamento de Fisica, Grupo de Fisica Teorica,
Universidad de Murcia,  Murcia, Spain.
}
\end{center}

\begin{abstract}
We present here a recopilation of recent results about
the possibility of detecting solar electron antineutrinos 
produced by solar core and convective magnetic fields.
These antineutrinos  are 
predicted by spin-flavor oscillations at a significant rate 
even if this mechanism is not the leading solution to the 
SNP. 
Using the recent Kamland results and 
assuming a concrete model for 
antineutrino production by spin-flavor precession in the 
convective zone based on chaotic magnetic fields,
we obtain bounds on the flux of solar antineutrinos,
on the average conversion neutrino-antineutrino probability
and on  intrinsic neutrino magnetic moment
In the most conservative case, 
$\mu\lsim 2.5\times 10^{-11}\ \mu_B$ (95\% CL).
When studying the effects of a core magnetic field,
we find in the weak limit a scaling of the antineutrino 
probability with respect to the magnetic field profile 
in the sense that the same probability function can be 
reproduced by any profile with a suitable peak field value. In 
this way the solar electron antineutrino spectrum can 
be unambiguosly predicted. We use  this scaling and the 
negative results indicated by the KamLAND experiment to 
obtain  upper bounds on the solar electron antineutrino flux.   
We find that, for a wide family of magnetic field profiles 
in the sun interior, the antineutrino appearance probability 
is largely determined by the magnetic field intensity but not 
by its shape. Explicit limits on neutrino transition 
moments are also obtained consistent with the convective 
case.  
These limits are therefor 
largerly independent of the detailed structure of the 
magnetic field in the solar interior.

\end{abstract}

\vfill

{\em Expanded version of the presentation 
contributed to 
  `` 8th International Workshop On 
Topics In Astroparticle And Underground Physics (TAUP 2003)''} 

%
%

\newpage

\renewcommand{\thefootnote}{\arabic{footnote}}
\setcounter{footnote}{0}

\section{Introduction}
Evidence of eelctron antineutrino disappearance in a beam of antineutrinos 
in the KamLAND experiment has been recently 
presented \cite{kloctober}.
The analysis of these results \cite{kloctober,klothers} 
in terms of neutrino oscillations
have   largely 
improved our knowledge of neutrino mixing in the LMA region.
The results appear to confirm in a independent way that the observed 
deficit of solar neutrinos is indeed due to neutrino oscillations. 
The ability to measure the LMA solution, the one
preferred by the solar neutrino data at present,
 ``in the lab'' puts KamLAND in a pioneering situation: after these 
results there should remain little doubt of the physical reality of 
neutrino mass and oscillations.
Once neutrino mass is observed,
neutrino magnetic moments are an inevitable consequence in the 
Standard Model and beyond. Magnetic moment interactions 
arise in any renormalizable gauge theory only as finite 
radiative corrections: the diagrams which contribute to the
neutrino mass will inevitably generate a magnetic moment
once the external photon line is added.

The spin flavor precession (SFP) \cite{SV,LM,Ak},
based on the interaction of the neutrino magnetic 
moment with the solar magnetic field was, second to oscillations, the most 
attractive scenario \cite{generalrandom}.    
SFP, although certainly not playing the major role in the solar neutrino deficit,
may still be present as a subdominant process, provided neutrinos have a 
sizeable transition magnetic moment. Its signature will be the appearance of
solar antineutrinos \cite{LM,Ak1,alianiantinu} 
which result from the combined effect of the vacuum
mixing angle $\theta$ and the transition magnetic moment $\mu_{\nu}$ converting 
neutrinos into antineutrinos of a different flavor. 
This can be schematically 
shown as 

\be
{\nu_e}_L \rightarrow {\nu_{\mu_L}} \rightarrow {\bar\nu_{e_R}},
\ee
\be
{\nu_e}_L \rightarrow {\bar\nu_{\mu_R}} \rightarrow {\bar\nu_{e_R}}
\ee
with oscillations acting first and SFP second in sequence (1) and in reverse
order in sequence (2). Oscillations and SFP can either take place in the same 
spatial region, or be spatially separated. Independently of their origin,
antineutrinos with energies above 1.8 MeV can be detected in KamLAND via the
observation of positrons from the inverse $\beta$-decay reaction 
$\bar\nu_{e} + p \rightarrow n+e^{+}$ and must all be originated from 
$^8 B$ neutrinos.

The KamLAND experiment is the 
successor of previous reactor experiments (
CHOOZ~\cite{CHOOZ}, PaloVerde~\cite{PaloVerde}) 
at a much larger scale in 
terms of baseline distance and total incident flux.  
This experiment relies upon a 1 kton liquid scintillator
 detector   located at the old, enlarged,  Kamiokande site.
 It searches for the oscillation of antineutrinos 
emitted by several nuclear power plants in Japan. 
The nearby 16  (of a total of 51) nuclear power stations deliver 
a $\overline{\nu}_e$ flux of $1.3\times 10^6 cm^{-2}s^{-1}$
for neutrino energies $E_\nu>1.8$ MeV at the detector position. 
About $85\%$ of this flux comes from  reactors forming a 
well defined baseline of 139-344 km. Thus, the flight range 
is limited  in spite of using  several reactors, because of this 
fact the sensitivity of KamLAND   increases by nearly two 
orders of magnitude compared to previous reactor experiments.

Beyond reactor neutrino measurements, the  secondary 
physics program 
of KamLAND includes diverse objectives as the measurement of geoneutrino flux 
emitted by the radioactivity  of the earth's crust and mantle, the detection 
of antineutrino bursts from galactic supernova and, after extensive  
improvement of the detection sensitivity, the detection of low energy 
${}^7 Be$ neutrinos using neutrino-electron elastic scattering.

Moreover, the KamLAND experiment is  capable of 
detecting  potential electron antineutrinos produced on fly 
from solar ${}^8$B neutrinos \cite{alianiantinu}. 
These  antineutrinos are 
predicted by spin-flavor oscillations at a significant rate if the neutrino is a Majorana particle and 
if its magnetic moment is high enough
\cite{generalrandom,specificrandom}.  
In Ref.\cite{alianiantinu} as been remarked 
 that the flux of reactor antineutrinos at the Kamiokande site 
is comparable, and in fact smaller, to the flux of 
${}^8$B neutrinos emitted by the sun
,$\Phi( {}^8 B)\simeq 5.6\times 10^6 cm^{-2}s^{-1}$ 
\cite{kloctober,bpb2001,Ahmad:2002jz}.
Their energy spectrum  is important at energies 
$2-4$ MeV while solar neutrino spectrum peaks at around 
$9-10$ MeV. As the inverse beta decay reaction cross section
increases as the square of the energy, we would 
expect nearly 10 times 
more solar electron antineutrino events even if the 
 initial fluxes were equal in magnitude.

The publication of the  SNO 
results~\cite{Ahmad:2002ka,Ahmad:2002jz}
has already 
made an important breakthrough towards the solution of the long standing
 solar neutrino 
\cite{Aliani:2002ma,Strumia:2002rv,Aliani:2002er,Aliani:2001zi}
problem (SNP) possible.
These results provide the strongest evidence so 
far (at least until KamLAND 
improves its statistics) for flavor oscillation 
in the neutral lepton sector.

The existing  bounds on solar electron antineutrinos are 
 strict.
 The present upper limit on the absolute flux of solar antineutrinos
originated from ${}^8 B$ neutrinos is 
\cite{alianiantinu,antibounds,PDG2002}
$\Phi_{\overline{\nu}}({}^8 B)< 1.8\times 10^5\ cm^{-2}\ s^{-1}$
which is equivalent to an averaged conversion probability bound
of $P<3.5\%$ (SSM-BP98 model). There are also bounds on their differential 
energy spectrum \cite{antibounds}: the conversion probability is smaller 
than $8\%$ for all $E_{e,vis}>6.5$ MeV going down the $5\%$ level above 
$E_{e,vis}\simeq 10$ MeV.

The main aim of this work is to study the implications of the 
recent KamLAND results on the determination of the 
solar electron antineutrino appearance probability, independently from 
concrete models on antineutrino production. 
We obtain upper limits on the solar antineutrino flux, the intrinsic magnetic moment
and the magnetic field at the bottom of the convective zone were 
 \cite{alianiantinu} from the published KamLAND data. 
In the second part of the work, we address
however a different antineutrino production model where the magnetic field 
at the solar core is the relevant one. 
The purpose of this part is to relate the solar magnetic field profile to
the solar antineutrino event rate in KamLAND which is a component of 
the total positron 
event rate in the  reaction above. In a previous paper \cite{Akhmedov:2002mf}
the question of what can be learned about the strength and coordinate
dependence of the solar magnetic field in relation to the current upper 
limits on the solar $\bar\nu_{e}$ flux was addressed. The system of equations
describing neutrino evolution in the sun was solved analytically in perturbation 
theory for small $\mu_{\nu}{B}$, the product of the neutrino magnetic moment
by the solar field. The three oscillation scenarios with the best fits were 
considered, namely LMA, LOW and vacuum solutions. In particular for LMA it was 
found that the antineutrino probability depends only on the magnitude of the 
magnetic field in the neutrino production zone. Neutrinos were, in the 
approximation used, considered to be all produced at the same point 
($x=0.05R_S$), where $^8 B$ neutrino production is peaked. 
In this work we will consider the more realistic case of a convolution 
of the production distribution spectrum with the field profile in that region.
It will be seen that this convolution leads to an insensitiveness of the
antineutrino probability with respect to the solar magnetic field profile,
in the sense that different profiles can correspond to the same probability
function, provided the peak field values are conveniently scaled. As a
consequence, an upper bound on the solar antineutrino flux can be derived which
is independent of the field profile and the energy spectrum of this flux
will also be seen to be profile independent.

The structure of this work is the following.
In section 2
 we discuss the main features of KamLAND experiment 
that are relevant for our analysis:
The
salient aspects of the procedure we are adopting and 
the  results of our analysis are presented and discussed in
 sections 3. 
In Section 4 we apply the results we obtained in a particular 
model for the solar magnetic field, we obtain bounds on the 
values of the intrinsic neutrino transition magnetic moments.
Finally, 
in section 5 we draw our 
conclusions and discuss possible future scenarios.

\vspace{2cm}

\begin{figure}[h]
\centering
\begin{tabular}{c}
\psfig{file=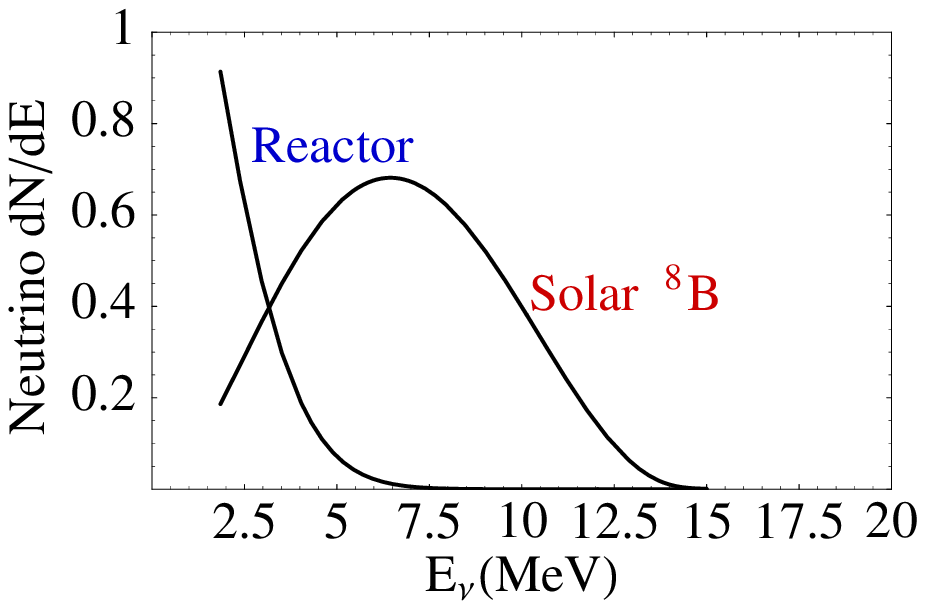,width=9cm} \\ 
\psfig{file=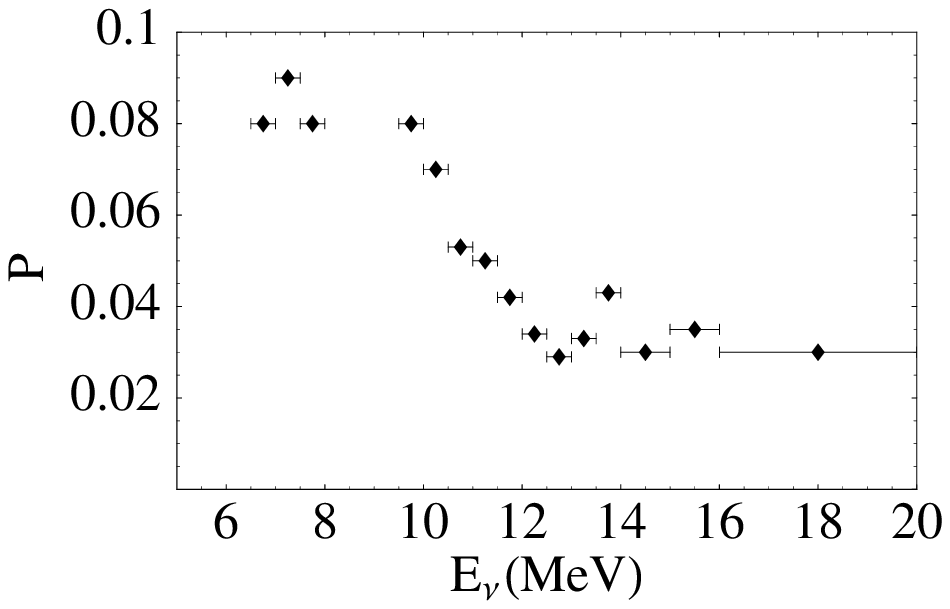,width=9cm}
\end{tabular} 
\caption{(Top)
The reactor antineutrino  and solar ${}^8 B$ neutrino 
\protect\cite{bpb2001} fluxes.
(Bottom).
Upper limits on solar antineutrino conversion probabilities 
(from Ref.\protect\cite{antibounds}).
}
\label{figurebounds}
\end{figure}

\clearpage

\section{A KamLAND overview}
\label{kamland}

Independently of their origin, solar or reactor
electron antineutrinos from nuclear reactors  with energies 
above 1.8 MeV can be detected in  KamLAND by  the inverse 
$\beta$-decay reaction $\overline{\nu}_e+p\to n+e^+$. The time 
coincidence, the space correlation and the energy balance  
between the positron signal and the 2.2 MeV $\gamma$-ray
 produced by the capture of a already-thermalized  neutron on a
 free proton make it possible to identify this reaction 
unambiguously, even in the presence of a rather large background. 

The main  ingredients in the calculation of the 
corresponding expected 
signals in KamLAND are  solar fluxes mentioned above, the reactor  flux and
the antineutrino cross section on protons. These last two 
 are considered below (see also Ref.\cite{Aliani:2002ca}). 

\subsection{The reactor antineutrino  flux}

We first describe the flux of antineutrinos coming from 
the power reactors.
A number of short baseline experiments 
(Ref.\cite{Murayama:2000iq} and references therein) 
have measured the energy spectrum of reactors at distances 
where oscillatory effects have been shown to be inexistent. 
They have shown that the theoretical neutrino flux predictions 
are reliable within 2\% \cite{piepke}.

The effective flux of antineutrinos released by the nuclear
 plants is a rather well  understood function of 
the thermal power of the reactor and
 the amount of thermal power emitted during the 
fission of a given nucleus, which gives the total amount, and 
the  isotopic composition of the reactor fuel which gives the 
spectral shape.
Detailed tables for these
 magnitudes can be found in Ref.~\cite{Murayama:2000iq}.

For a given isotope $(j)$ the energy  spectrum can be parametrized 
by the following expression 
$d N_\nu^{j}/d E_\nu=\exp (a_0+a_1 E_\nu+ a_2 E_\nu^2)$
where the coefficients $a_i$ depend 
on the nature of the fissionable isotope 
(see Ref.\cite{Murayama:2000iq} for explicit values).
Along the year, between periods of refueling, the total effective flux changes with time as the fuel is expended and the isotope 
relative composition varies.
The overall spectrum is at a given time
$ {d N_\nu}/{d E_\nu}=\sum_{j=isotopes} 
c_j(t){d N_\nu^{j}}/{d E_\nu}.$
To compute a fuel-cycle averaged spectrum
we have made use of the typical time 
evolution of the relative 
abundances $c_j$, which  can be seen in Fig. 2 of 
Ref.\cite{Murayama:2000iq}.
This averaged spectrum can  be again  fitted very well by 
the same functional expression as above.
The isotopic energy yield  is properly taken into account. 
As the result of this fit, we obtain 
the following values which are the ones to be used in the 
rest of this work:
$a_0=0.916,\quad a_1=-0.202,\quad a_2=-0.088.$   
Although individual variations of the $c_j$ along the 
fuel cycle can be very high, the variation of the two most 
important ones is highly correlated: the 
coefficient $c({}^{235} U)$ increases in the range
 $\sim 0.5-0.7$ while 
$c({}^{239} Pu)$ decreases $\sim 0.4-0.2$. 
This correlation makes  the effective description of the 
total spectrum by a single expression as above useful.
With the fitted coefficients $a_i$ above, the difference between
this effective spectrum and the real one is typically $2-4\%$ 
along the yearly fuel cycle.

\subsection{Antineutrino cross sections}

We now consider the
cross sections for antineutrinos on protons. We will 
sketch the form of the well known differential expression and 
more importantly we will give updated numerical values 
for the transition matrix elements which appear as 
coefficients. 

In the limit of infinite nucleon mass, the 
cross section  for the reaction 
\anureaction\  is given by \cite{zacek,reines} 
$\sigma(E_{\overline{\nu}})=k\  E_{e^+} p_{e^+}$
where $E,p$ are the  positron energy and momentum
 and $k$ a  transition matrix element which will be 
considered 
below.
The  positron spectrum  is  monoenergetic: 
 $E_{\overline{\nu}}$ and  
$E_{e^+}$ are related by:
$E_{\overline{\nu}}^{(0)}=E_{e^+}^{(0)}+\Delta M$,
where $M_n, M_p$ are the neutron and proton masses
 and $\Delta M=M_n-M_p\simeq 1.293 $ MeV.

Nucleon recoil corrections are 
potentially important in relating the positron and antineutrino 
energies in order to evaluate the antineutrino flux. 
Because the antineutrino 
flux $\Phi(E_\nu)$ would typically decrease quite 
rapidly with energy, the lack of adequate corrections 
will systematically overestimate the   positron yield.   
For both cases, solar or reactor antineutrinos, 
 because the antineutrino 
flux $\Phi(E_\nu)$ would typically decrease quite 
rapidly with energy, the lack of adequate corrections 
will systematically overestimate the   positron yield.   
For the solar case and taking into account the SSM-BP98 
${}^8 B$ spectrum, the effect  decrease the positron 
yield by 2-8\% at the main visible energy range $\sim 6-10$ MeV.
The positron yield could decrease up 50\% at {\em hep} 
neutrino energies, a region where incertitudes in the total 
and differential spectrum are of comparable size or larger.
 Finite energy resolution smearing will however diminish  
this correction when integrating over large enough energy 
bins: in the range $6.5-20$ MeV the net positron suppression 
is estimated to be at the $5\%$ level, increasing up $20\%$ 
at {\em hep} energies.

At highest orders,  the  positron spectrum  is not 
monoenergetic and one has to integrate over the positron angular 
distribution to obtain the positron yield.
We have used the 
complete expressions which  can be found 
in Ref. \cite{vogel}. Here we only want to stress 
the numerical value of 
the overall coefficient $\sigma_0$ (notation of 
Ref.\cite{vogel}) which is related to the 
transition matrix element $k$ above.
The matrix transition element can be written in terms of 
measurable quantities as 
$k=2\pi^2 \log 2/ (m_e^5 f \, t_{1/2}).$
Where the value of the space factor
  $f=1.71465\pm 0.00015$ follows from calculation 
\cite{zacek21},
 while $t_{1/2}=613.9\pm 0.55$ sec is the latest published value for 
the free neutron half-life \cite{PDG2002}. This value has a  significantly 
smaller  error than previously quoted measurements.
From the values above, we obtain the extremely precise value: 
$k=(9.5305\pm 0.0085)\times 10^{-44}\  cm^2/MeV^{2}.$ 
From here the coefficient which appears in the differential 
cross section is obtained as
(vector and axial vector couplings $f=1,g=1.26$):
$ k=\sigma_0 (f^2+3 g^2).$
In summary, the differential cross section which appear in
  KamLAND are very well known, its theoretical 
errors are negligible if updated values are employed.

\section{The   solar signal and reactor backgrounds}

Electron antineutrinos from any source, nuclear reactors or
solar origin,  with energies 
above 1.8 MeV are measured in KamLAND by detecting the inverse 
$\beta$-decay reaction $\overline{\nu}_e+p\to n+e^+$. The time 
coincidence, the space correlation and the energy balance  
between the positron signal and the 2.2 MeV $\gamma$-ray
 produced by the capture of a already-thermalized  neutron on a
 free proton make it possible to identify this reaction 
unambiguously, even in the presence of a rather large background. 

The two principal ingredients in the calculation of the 
expected 
signal in KamLAND are  the corresponding   flux and
the electron antineutrino cross section on protons.
The average number of positrons $N_i$ originated from the 
solar source which 
are detected per visible energy bin $\Delta E_i$ is given by the convolution 
of different quantities: 
\begin{eqnarray} \hspace{-0.3cm}
N_i&=& Q_0 \int_{\Delta E_i}dE_e \int_0^\infty dE_e^r \epsilon(E_e)R(E_e,E_e^r)
\int_{E_e^r}^\infty dE_{\overline{\nu}} \overline{p}(E_{\overline{\nu}})
\Phi (E_{\overline{\nu}})
 \sigma (E_{\overline{\nu}},E_e^r)
\label{e3466} 
\end{eqnarray}
where $Q_0$ is a normalization constant  accounting for the 
fiducial volume and live time of the experiment,
$\overline{p}$.
Expressions for the 
electron antineutrino capture cross section  
$\sigma (E_{\overline{\nu}},E_e^r)$ 
are taken from 
the literature \cite{vogel,kltorrente}. The  matrix 
element  for this cross section can 
be written in terms of the neutron half-life, 
 we have used the latest published 
value $t_{1/2}=613.9\pm 0.55$ \cite{PDG2002}.
The functions $\epsilon(E_e)$ and $R(E_e,E_e^r)$ are
 the detection efficiency and  the energy 
resolution function.
We use in our analysis  the following
expression for the energy resolution in the prompt 
positron detection
$\sigma(E_e)=0.0062+ 0.065\surd E_e$ .
This expression is obtained from the raw calibration 
data presented in Ref.\cite{klstony}.
Note that we prefer to use this  expression instead of the 
much less accurate one given in 
Ref.\cite{kloctober}.
Moreover, we assume a  408 ton fiducial 
mass and  the detection efficiency is 
taken  independent of the energy \cite{kloctober},
 $\epsilon=80\%$.
In order to obtain concrete limits, a model should  be taken which 
predict $\overline{p}$ and its dependence with the energy.
For our purpose it will suffice  to suppose  $\overline{p}$ a 
constant. This is justified at least in two cases: a) 
if the energy range $\Delta E$ over which we perfom the integration 
is small enough 
so the variation of the probability is not very large, or b) if 
we reinterpret ${\overline p}$ as an energy-averaged probability,
 note that, in a general case, this is always true because the 
un-avoidable convolution with a  finite energy resolution. 
(see Expression 10 in Ref.\cite{specificrandom}):
\begin{eqnarray}
{\overline p}_{\Delta E}&=&  
\int_{\Delta E} dE\ \sigma(E) \Phi(E) P_{\overline \nu}(E)/
{\int_{\Delta E} dE\ \sigma(E) \Phi(E)}.
\label{avnu}
\end{eqnarray}
Let us finally note that independently of the reasons above, 
upper 
limits to be obtained on continuation are still 
valid even if the antineutrino probabilities are significantly 
different from constant: if we take 
${\overline p}=\max_{\Delta E} P_{\overline \nu}(E)$ 
the  expected antineutrino signal computed 
with ${\overline p}$ will 
be always larger than the signal obtained inserting the full 
probability.

Similarly,  the expected numbers of positron events originated 
from power reactor neutrinos are obtained summing 
the expectations for all the relevant 
reactor sources weighting each source by its power and distance to the detector
(table II in Ref.~\cite{Murayama:2000iq}), assuming the same spectrum originated 
from each reactor. 
We have used the antineutrino flux spectrum  given by  
the expression  of the previous section and  the 
relative reactor-reactor power normalization.

For one year of running with the 600 ton fiducial mass and for standard nuclear 
plant power and fuel schedule: we assume all the reactors operated at 
$\sim 80\%$ of their maximum capacity and an averaged, time-independent, fuel 
composition equal for each detector, the experiment expects about 550 
antineutrino events.

In addition to the  reactor antineutrino  signal deposited in the detector, 
two classes of other backgrounds can be distinguished 
\cite{piepke,Murayama:2000iq,brackeeler}. The so called random coincidence 
background is due to the contamination of the detector scintillator by 
U, Th and Rn. From MC studies and assuming that an adequate level of
purification can be obtained, the background coming from this source
is expected to be $\sim 0.15$ events/d/kt which is 
equivalent to a signal to background ratio of  $\sim 1\%$. 
Other works \cite{usareport} conservatively estimate 
a $5\%$ level for this ratio. More importantly for what it follows, 
one expects that the 
random coincidence  backgrounds will be a relatively steeply 
falling function of energy. The assumption of no random coincidence 
background should be relatively safe at high energies above $\sim 5$ MeV
which are those of interest here.

The second source of background, the so called correlated background is 
dominantly caused by cosmic ray muons and neutrons. The KamLAND's depth is
 the main tool to suppress those backgrounds. MC methods estimate a correlated 
background of around $0.05$ events/day/kt distributed over all the energy 
range up to $\sim 20$ MeV, this is the quantity that we will consider later.


In order to estimate the sensitivity of KamLAND to 
put limits on the flux of antineutrinos arriving from the 
sun we have computed the expected signals coming from solar and 
reactor  antineutrinos and from the background. 
They  are presented in Table (\ref{table1})
 for different representative 
values of the minimum energy required ($E_{thr}$) for the 
visible positrons. 
We have supposed a background of $0.05$ evt/d/kt uniformly 
 distributed over the full energy range.
To obtain the solar numbers (first column, $S_{sun}$)
 we have supposed full  neutrino-antineutrino conversion  
($\overline{P}=1$) 
with no spectral distortion.
For any other conversion probability, the experiment should see 
the  antineutrino 
quantity $\overline{P}\times S_{sun}$ 
in addition to the reactor ones and other background. 
If the experiment does not receive any solar antineutrinos, making 
a simple statistical estimation (only statistical errors are 
included) we obtain the upper limits on the 
conversion probability which appear in the last column of the table.

From the table we see that 
after three years of data taking the optimal result is obtained imposing a 
energy detection threshold at $\sim 7$ MeV. A negative result would allow 
to impose an upper limit on the average antineutrino appearance probability 
at $\sim 0.20\%$ (95\% CL). The corresponding limits after one year of data 
taking are only  slightly worse, they are respectively: 0.21-0.24\% (95\% CL).

\begin{figure}[h]
\centering
\psfig{file=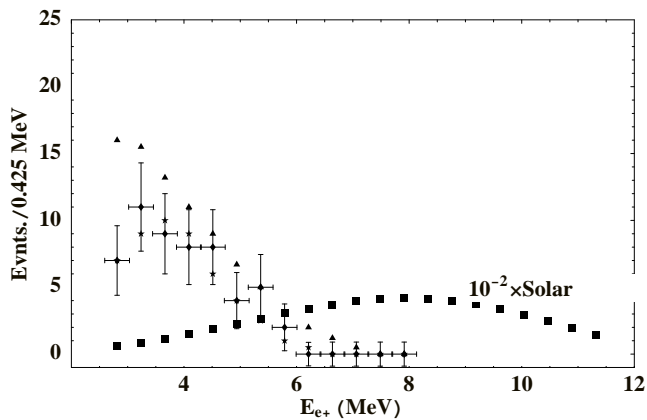,width=9cm} 
\label{f1}
\caption{\small
The KamLAND  positron spectrum from reactor antineutrinos 
(from Fig.5 in Ref.\protect\cite{kloctober}):
 measured (145.5 days), MC expectations in absence of oscillations
and best fit including neutrino oscillations 
($\Delta m^2=6.9\times 10^{-5}$ eV$^2$, $\sin^2\theta=1$,
respectively points with error-bars, triangles and stars).
The ``solar'' positron spectrum  (black solid squares) 
obtained assuming the 
shape of the ${}^8 B$ 
neutrino flux and a total normalization
$10^{-2}\times \Phi({}^8 B)$ (that is, an overall 
$\nu_e-\overline{\nu}_e$ 
conversion probability $\overline{P}\sim 1\%$).}
\end{figure}

The results of our simulation are summarized
 in Fig.\ref{f1} where we show  
the ``solar'' positron spectrum  obtained assuming the 
shape of the ${}^8 B$ 
neutrino flux and a total normalization
$10^{-2}\times \Phi({}^8 B)$ which means an overall 
$\nu_e-\overline{\nu}_e$ 
conversion probability $\overline{P}\sim 1\%$.

These results are obtained under the supposition of no disappearance on 
the reactor flux arriving to KamLAND. No flux suppression is  expected for 
values of the mixing parameters in the LOW region, more precisely for any
$\Delta m^2 \leq 2 \times 10^{-5}$ eV$^2$ (see Plot 1(right) in 
Ref.\cite{Aliani:2002ma} and Ref.\cite{Aliani:2002ca}).
The consideration of  reactor antineutrino oscillations does not change 
significantly the sensitivity in obtaining upper limits on $\overline{P}$.
For values of the mixing parameters fully on the LMA region,
$\Delta m^2 \geq 1-9 \times 10^{-5}$ eV$^2$, the flux suppression is 
typically $S/S_0\sim 0.5-0.9$ and always over $S/S_0\sim 0.4$, for any 
the energy threshold $E_{thr}\sim 5-8$ MeV.
We have obtained the expected reactor antineutrino contribution  
for a variety of points in the LMA region 
(see table I in Ref.\cite{Aliani:2002ca}) and 
corresponding upper limits on $\overline{P}$: the results after 
3 years of running are practically the same while for 1 year of 
data running are slightly better (for example $\overline{P}$ goes 
down from 0.27 to 0.3 for $E_{thr}>6$ MeV. 

\clearpage

\phantom{ppp}

\vspace{2cm}

\begin{table}[h]
 \begin{center}
  \scalebox{0.99}{
    \begin{tabular}{lcrccc}
\hline
  E$_{thr}$     & $S_{Sun}$& $S_{Rct}$ & Bckg. &
 P (CL 95)\% & P (CL 99)\% \\
\hline
6 MeV  & 616  & 43 & 70 & 0.22 & 0.23          \\
7 MeV  & 500  & 11 & 65 & 0.19 & 0.20         \\
8 MeV  & 366  & 2 &  60 & 0.21 & 0.23         \\
\hline
      \end{tabular}
}
 \caption{\small Expected signals from solar antineutrinos  after 
3 years of data taking. Reactor antineutrino (no oscillation is 
assumed) and other background 
(correlated background) over the same period. The random coincidence 
background is supposed negligble above these energy thresholds. 
Upper limits on the antineutrino oscillation probability.
}
  \label{table1}
 \end{center}
\end{table}

\vspace{2cm}

\clearpage

\section{Analysis and Results}

We will obtain upper bounds on the solar electron antineutrino 
appearance probability analyzing the observed KamLAND rates
in three different ways (see Refs.\cite{torrentemilan,alianiantinu}, see also Refs.\cite{torrentepulido}). In the first one, we will make a 
standard $\chi^2$ analysis of the observed and expected solar 
signals in the 13 prompt positron energy bins considered by 
KamLAND \cite{kloctober}. 
In the second and third cases we will 
apply Gaussian and poissonian probabilistic 
considerations to the 
global rate seen by the experiment and to the individual 
event content in the
highest energy bins ($E_e> 6 $ MeV) where KamLAND 
observes zero events.
This null signal 
 makes particularly simple the extraction of statistical 
conclusions in this case.

{\bf  Analysis of the KamLAND Energy Spectrum }

Here we fully 
use the  binned KamLAND signal (see  Fig. 5 in  Ref.\cite{kloctober}) 
for  estimating
 the parameters of solar electron antineutrino production 
from the method of  maximum-likelihood.
We minimize the quantity
$\chi^2 = \chi^2_{i=1,9}+\chi^2_{i=10,13}$
where the first term correspond to the contribution of the
first nine bins where the signal is large enough and  
 the use of the
 Gaussian approximation is justified. 
The second term correspond
to the latest bins where the observed and expected signals
are very small and poissonian statistics is needed. 
The explicit expressions are:
\begin{eqnarray}
\chi^2_{i=1,9}&=& \sum_{i=1,9} 
\frac{(S_i^{exp}-S_i^{teo})^2}{\sigma^2}   \\
\chi^2_{i=10,13} &=&2\sum_{i=10,13} 
 S_i^{teo}-S_i^{exp}+ S_i^{exp}\log \frac{S_i^{exp}}{S_i^{teo}}.
\end{eqnarray}
The quantities $S_i$ are the observed bin contents from KamLAND.
The theoretical signals are in principle a function 
of three different parameters: the solar electron antineutrino 
appearance probability $\overline{p}$  and
 the neutrino oscillation parameters $(\Delta m^2,\theta)$.
Both contributions, the contribution from solar antineutrinos 
and that one from solar reactors, can be treated as different
summands:
\begin{eqnarray}
S_i(\overline{p},
\Delta m^2,\theta)&=&
S_i^{solar}(\overline{p})+
S_i^{reactor}(\Delta m^2,\theta).
\end{eqnarray}
According to our model, 
the solar antineutrino 
appearance probability  $\overline{p}$ is  
taken as a constant and we can finally write:
\begin{eqnarray}
S_i(\overline{p},
\Delta m^2,\theta)&=&
\overline{p}\times S_i^{0}+
S_i^{reactor}(\Delta m^2,\theta).
\end{eqnarray}
In this work
we will take for the minimization 
values of the oscillation parameters those obtained 
when ignoring any solar antineutrinos
(LMA solution 
$\Delta m^2=6.9\times 10^{-5}$ eV$^2$, $\sin^2\theta=1$
 from Ref.\cite{kloctober}) and 
we will perform a one-parameter minimization with respect
$\overline{p}$. 
This approximation is well justified because the solar 
antineutrino probability is clearly very small, 
We avoid in this way  the simultaneous 
 minimization 
 with respect to the three parameters 
($\overline{p},  \Delta m^2,\theta)$.

We perform a 
 minimization of the one dimensional  function
 $\chi^2(\overline{p})$. 
to test a particular oscillation hypothesis against the 
parameters of the best fit 
and obtain the allowed interval  in 
$\overline{p}$
parameter space 
taking into account the asymptotic properties of the 
likelihood function, i.e. $\log {\cal L}-\log {\cal L}_{min}$
behaves asymptotically as a $\chi^2$ with one degree of 
freedom.
In our case, the  minimization can be performed analytically because 
of the simple, lineal, dependence. 
A given point in the confidence interval
 is allowed if 
 the globally subtracted quantity 
fulfills the condition 
 $\Delta\chi^2=
\chi^2 (\overline{p})-\chi_{\rm min}^2<\chi^2_n(CL)$.
Where $\chi^2_{n=1}(90\%,95\%,...)=2.70,3.84,..$ are the 
quantiles for one  degree of freedom.

Restricting to physical values 
of $\overline{p}$, the minimum 
of the $\chi^2$ function is obtained for 
$\overline{p}=0$. The corresponding confidence intervals are
$\overline{p}<4.5 \% $ (90\% CL) and 
$\overline{p}<7.0\% $ (95\% CL). 
We have explicitly checked, varying the concrete place
where the division 
between ``Gaussian'' and ``poissonian'' bins is 
established 
that the values of these upper limits are 
largely insensitive to details of our analysis.
In particular, similar upper limits are obtained
in the extreme cases: if  Gaussian or poissonian 
statistics is employed for all 13 bins.
These upper limits are considerably weaker than those 
obtained in the next section. One possible  reason for that is that 
they are  obtained applying asymptotic general arguments to the 
$\chi^2$ distribution, stronger, or more precise  
limits could be obtained if a  
Monte Carlo simulation of the distribution of the 
finite sample $\chi^2$ distribution is performed 
(where the boundary condition $\overline{p}>0$ should 
be properly included).

{\bf Analysis of the global rate and highest energy bins}

We can make an estimation of the upper bound on the 
appearance solar electron antineutrino probability simply counting 
the number of observed events and subtracting the 
number of events expected from the best-fit oscillation 
solution. For our purposes this difference, which in this 
case is positive, can be interpreted as a hypothetical 
signal coming from  
solar antineutrinos
 ($\Delta m^2_0=6.9\times 10^5\ eV^2,
\sin^2 \theta_0=1)$.:
\begin{eqnarray}
S_{solar}=\overline{p}\times S_{solar}^0&=& 
S_{obs}-S_{react}(\Delta m^2_0,\sin^2 \theta_0).
\end{eqnarray}
Putting \cite{kloctober} 
$S_{obs}=54.3\pm 7.5 $ and
$S_{react}(\Delta m^2_0,\sin^2 \theta_0)= 49\pm 1.3 $,
 we obtain
$S_{obs}-S_{react}< 64.8\  (67.2)$ at 90 (95)\% CL.
From these numbers, the corresponding limits on 
solar electron antineutrino appearance probability are
$\overline{p}< 0.45\%, \ 0.52\% $ at 90 or 95\% CL. 
These limits are valid for the neutrino energy range
 $E_{\nu}\sim 2-8 $ MeV. 
In this case, due to the large range, the limits are 
better interpreted as limits on an 
energy-averaged probability
 according to expression \ref{avnu}. 

In a similar approach, we use on continuation 
the  binned KamLAND signal corresponding to 
the four highest energy bins
(see  Fig.\ref{f1}) which, as we will see, provide the strongest 
statistical significance and bounds. The reason for that is   
that the experiment KamLAND does not observe any signal here 
and, furthermore,
 the expected signal from oscillating neutrinos with LMA 
parameters is negligibly small.
Due to the small sample,
we apply Poisson statistics to any of these bins and use 
the fact that a sum of Poisson variables of mean $\mu_i$ is 
itself a Poisson variable of mean $\sum \mu_i$.
The background (here the reactor antineutrinos) and the 
signal (solar electron antineutrinos) are assumed  to be independent
Poisson Random variables with known means. 
If no events are 
observed, and, in particular, no background is observed, the 
unified intervals \cite{cousins,PDG2002}
$[0,\epsilon_{CL}]$ 
 are 
$[0,2.44]$ 
at 90\% CL and 
$[0,3.09]$ at 95\% CL.
From here, we obtain 
$\overline{p}\times S_0^{solar} < \epsilon_{CL}$ or 
$\overline{p}  < \epsilon_{CL}/S_0^{solar}$. Taking the 
expected number of events in the first 145 days of data taking 
and in this energy range (6-8 MeV) we obtain:
$\overline{p}<0.12 \% $ (90\% CL) and 
$\overline{p}<0.15\% $ (95\% CL).

\clearpage

\section{ A model for solar antineutrino production in the
sun convection zone}

\begin{figure}[h]
\centering
\psfig{file=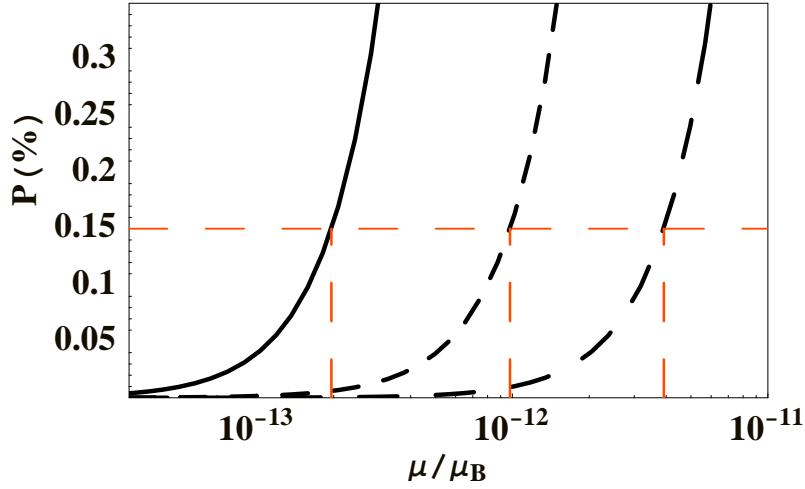,width=11cm} 
\caption{\small The solar antineutrino appearance probability $\overline{p}$ 
as a function of the transition neutrino magnetic moment, in units of 
Bohr magnetons $\mu_B$, for fixed values of the r.m.s solar magnetic 
field (Formula 
\protect\ref{anuprob}). From left (solid) to right (dashed), 
curves correspond to 
$B=1000,200,50$ kG. From the curves, an upper 
limit $\overline{p}<0.15\%$ implies 
$\mu< 1.9\times 10^{-13} \mu_B, 
9.0\times 10^{-13} \mu_B, 
3.0\times 10^{-12} \mu_B$ respectively for each of the magnetic 
field above. 
}
\label{f2}
\end{figure}

The combined action of  spin flavor precession in a 
magnetic field and  ordinary neutrino matter oscillations 
can produce an observable flux of $\anueR$'s from the Sun 
in the case of the neutrino being  a Majorana particle.
In the simplest model, where a thin layer of highly chaotic of 
magnetic field is assumed at the bottom of the convective 
zone (situated at $R\sim 0.7 R_\odot$), 
the antineutrino appearance 
 probability at the exit of the 
layer $P(\overline{\nu})$  is basically equal to 
the appearance probability of antineutrinos 
at the earth \cite{specificrandom,generalrandom} ( see also 
Refs.\cite{pulido} for some recent studies on RSFP solutions to
the Solar Neutrino Problem). 
The quantity  $P(\overline{\nu})$ is in general a function 
of the neutrino oscillation parameters  
$(\Delta m^2,\theta)$, the neutrino intrinsic magnetic moment
and also of the neutrino energy and 
the characteristics and magnitude of the solar 
magnetic field. However, in a accurate enough approximation, 
such probability can be factorized in a term depending 
only on the oscillation parameters and another one depending 
only on the spin-flavor precession parameters: 
\begin{equation}
P(\overline{\nu} )  = 
 \frac{1}{2}P_{e\mu}(\Delta m^2,\theta)\times 
\left [1-\exp\left (-4 \Omega^2 \Delta r  \right )  \right ]  
\label{anuprob}
\end{equation}
where $P_{e\mu}$ is the $e-\mu$ solar 
conversion probability. We will assume in this work the 
LMA central values for $(\Delta m^2,\theta)$ obtained 
from recent KamLAND data and which are compatible 
with the SNO observations in solar neutrinos \cite{alianidecember}, we will take 
$P_{e\mu}(\Delta m^2,\theta)\simeq
\langle P_{e\mu}\rangle^{exp,SNO} \simeq  0.4$. 
The second factor appearing in the expression 
contains  the 
effect of the magnetic field. 
This quantity depends on  the layer
width $\Delta r$  ($\sim 0.1 R_\odot$) and 
$\Omega^2\ \equiv  \frac{1}{3} L_0 \mu^2 \langle B^2\rangle$,
where
$\langle B^2\rangle$ 
the r.m.s strength of the
magnetic field 
and $L_0$ is a scale length ($L_0\sim 1000$ km).
For small values of the argument we have the following 
approximate expression 
which is accurate enough for many applications
$$P(\overline{\nu}) \simeq P_{e\mu}\times 2 \Omega^2 \Delta r=\kappa\ \mu^2 \langle B^2\rangle$$
the solar astrophysical  factor 
$\kappa\equiv 2/3 P_{e\mu} L_0 \Delta r$ is numerically 
$\kappa^{LMA}\simeq 2.8\times 10^{-44}$ MeV$^{-2}$.
Upper limits on the antineutrino appearance probability 
can be translated into upper limits 
 on the neutrino transition magnetic moment and the magnitude 
of the magnetic field in the solar interior.
The results of the Formula \ref{anuprob} 
can be seen in Figure \ref{f2}.
An upper  bound $\overline{p}< 0.15-0.20\%$ (95\% CL)
implies an upper limit on 
the product of the intrinsic neutrino magnetic moment and
the value of the convective solar magnetic field as 
$\mu B< 2.3\times  10^{-21}$ MeV (95\% CL).
In Fig.\ref{f2} we show the antineutrino probability 
as a function 
of the magnetic moment $\mu$ for fixed values of 
the magnitude of 
the magnetic field.
For realistic values of other astrophysical solar
 parameters ($L_0\sim 1000 $ km, $\Delta r\sim 0.1\ R_\odot$), 
these upper 
limits would imply that the neutrino magnetic 
moment is constrained to be, in the most desfavourable case, 
$\mu\lsim 3.9\times 10^{-12}\ \mu_B$ (95\% CL) for a relatively small
field $B= 50$ kG.  
Stronger limits are obtained for slightly 
higher values of the magnetic field:
 $\mu\lsim 9.0\times 10^{-13}\ \mu_B$ (95\% CL) for 
field $B= 200$ kG and 
 $\mu\lsim 2.0\times 10^{-13}\ \mu_B$ (95\% CL) for 
field $B= 1000$ kG. Let us note that these assumed values for 
the  magnetic field at the base the solar convective zone 
are relatively mild and well within present astrophysical 
expectatives.

\clearpage

\section{The magnetic field in the sun core}

\subsection{The solar antineutrino probability in core conversions}

We start (see Ref.\cite{torrentepulido}) with the probability that a ${\nu_e}_L$ 
produced inside the sun will reach the earth as a $\bar\nu_{e_R}$
\be
P(\nu_{e_L} \rightarrow \bar\nu_{e_R})=P(\nu_{e_L} \rightarrow 
\bar\nu_{\mu_R};R_S) \times 
P(\bar\nu_{\mu_R} \rightarrow \bar\nu_{e_R};R_{es})
\ee
in which the first term is the SFP probability, $R_S$ is the solar radius
and the second term is given 
by the well known formula for vacuum oscillations
\be
P(\bar\nu_{\mu_R} \rightarrow \bar\nu_{e_R};R_{es})={\sin}^{2}2\theta 
~~{\sin}^{2}\! \left(\frac{\Delta m^{2}}{4E}R_{es}\right)=\frac{1}{2}.
\ee
Here $R_{es}$ is the distance between the sun and the earth and the rest of
the notation is standard. Since $1.8 MeV < E < 15 MeV$ and, for LMA, 
$\Delta m^{2}=6.9\times10^{-5}eV^{2}$, $\sin^2 2\theta=1$ \cite {Eguchi:2002dm}, 
we take the $\bar\nu_{\mu_R} \rightarrow \bar\nu_{e_R}$ vacuum oscillations 
to be in the averaging regime.

The SFP amplitude in perturbation theory for small $\mu B$ 
is \cite{Akhmedov:2002mf} \footnote{For notation we refer the reader to 
ref. \cite{Akhmedov:2002mf}.}
\be
A(\nu_{e_L} \rightarrow \bar\nu_{\mu_R})= \frac{\mu B(r_i) \sin^{2}\theta (r_i)}{g^{'}_2(r_i)}.
\ee
A key observation is that the antineutrino appearance probability is 
dependent on the production point of its parent neutrino 
so that the overall antineutrino probability is
\be
P(\nu_{e_L} \rightarrow \bar\nu_{e_R})= \frac{1}{2} \int |A(\nu_{e_L} 
\rightarrow \bar\nu_{\mu_R})|^2 f_{B}(r_i)dr_i
\ee
where $f_{B}$ represents the neutrino production distribution function 
for Boron
neutrinos \cite{Bahc} and the integral extends over the whole production 
region. As shall be seen, owing to this integration, the energy shape of 
probability (6) is largely insensitive to the magnetic field profile.

\begin{figure}[h]
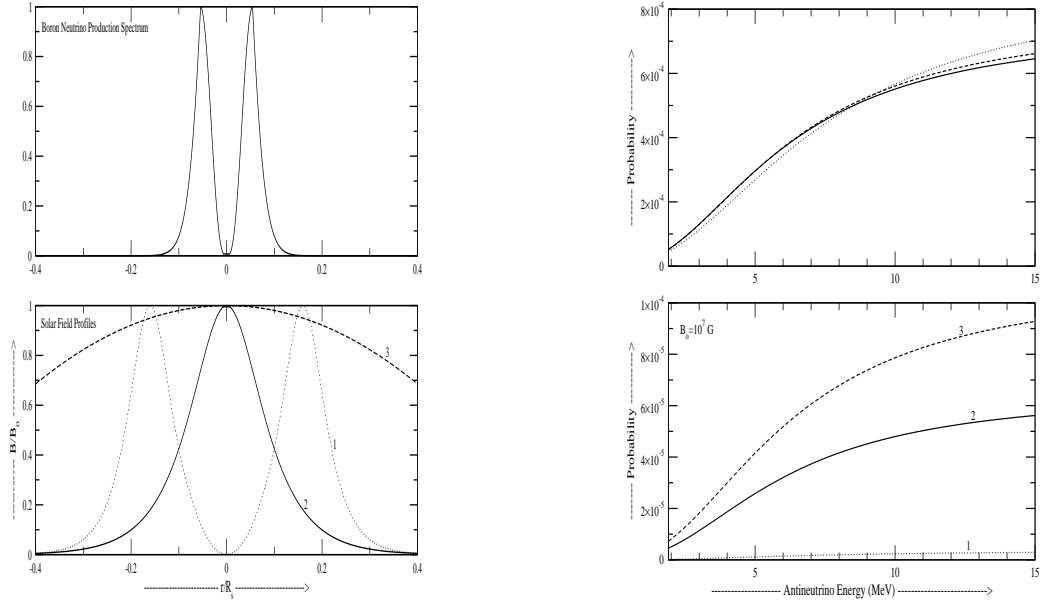

\begin{center}
\hspace*{-0.1cm}
\begin{tabular}{lr}
\epsfig{file=fin1.eps,height=8.0cm,width=5.5cm,angle=0} \hspace{2cm}&
\hspace{2cm}
\epsfig{file=fin2.eps,height=8.0cm,width=5.50cm,angle=0}
\end{tabular}
\end{center}
\caption{ \small 
(LEFT) Upper panel: $^8B$ neutrino production spectrum (in arbitrary 
units) as a function of the radial coordinate. Lower panel: the three solar field 
profiles considered in the main text normalized to $B_0$, the peak field value.  
(RIGHT) Antineutrino probabilities for solar field profiles 1, 2
and 3. Upper panel: the peak field is chosen in each case so as to produce 
the same event rate in KamLAND, (see the main text). Lower panel: the same
value of the peak field ($B_0=10^7G$) is seen in each case to lead to 
probabilities of quite different magnitudes.} 
\label{fig1}
\label{fig2}
\end{figure}

As mentioned above, for the LMA solution only the solar field profile in the
neutrino production region \cite{Akhmedov:2002mf} can affect the antineutrino 
flux. Hence
we will discuss three profiles which span a whole spectrum of possibilities at 
this region.
We study from a vanishing field (profile 1) to a maximum field at the solar center, with, 
in this second case, either a fast decreasing field intensity (profile 2) or a 
nearly flat one (profile 3) in the solar core 
(see fig. \ref{fig1}, lower panel). 
Thus, we consider 
respectively the following three profiles

{\it Profile 1}
\be 
B(r)=B_0[\cosh(9r)-1]~~,~~|r| \leq r_c
\ee 
\be
B(r)=B_0/\cosh[25(r-r_R)]~~,~~|r|>r_c,
\ee
with $r_c=0.08$, $r_R=0.16$,

{\it Profile 2}
\be
B(r)=B_0/\cosh(15r)~~,~~|r| \geq 0,
\ee

{\it Profile 3}
\be
B(r)=B_0[1-(r/r_c)^2]~~,~~|r| \leq r_c,
\ee
with $r_c=0.713$.

We
also show in fig. \ref{fig1} (upper panel) the $^8B$ production distribution spectrum,
so that a comparison between the strength of the field and the production
intensity can be directly made.

The antineutrino production probabilities as a function of energy for
each of these profiles are given in fig. \ref{fig2}. In the first panel,
the values of the peak field are chosen so as to produce a fixed number
of events. In this case the
probability curves differ only slightly in their shapes while their 
normalizations 
are the same. The curves are in any case similar to 
the SFP survival probability ones \cite{Pulido:1999xp} in 
the same energy range. 
In the second panel of fig. \ref{fig2} the antineutrino 
probabilities for a
common value of the peak field and these three different profiles are shown. 
It is hence apparent from these two graphs how the distribution of the 
magnetic field intensity is determinant for the magnitude of the 
antineutrino probability, but not for its shape. One important reason for 
this behavior is that we have integrated the antineutrino probability 
over the Boron production region.


\vspace{2cm}

\clearpage

\section{Results for the Magnetic Profile}

The antineutrino signal for any magnetic field profile $B(r)$
can be written, 
taking into account the previous formulas and the near invariance of the 
probability shape (see fig. \ref{fig2}), as 
\begin{eqnarray}
S_{\overline{\nu}}[{B(r)}]&=& \alpha S_{\overline{\nu}}^0
\end{eqnarray}
where $S_{\overline{\nu}}^0$ is the antineutrino signal taken at some 
nominal reference value $B_0^0$ for the field at the solar core for a certain 
reference profile $B^0$. This profile dependent parameter $\alpha$, being a ratio
of two event rates given by eq.(7) for different profiles, can thus be
simplified to 
\be
\alpha=\frac{\int \left(\frac{B(r_i)\sin^2\theta(r_i)}{g^{'}_{2}(r_i)}\right)^2 f_B (r_i)dr_i}
{\int \left(\frac{B^{0}(r_i)\sin^2\theta(r_i)}{g^{'}_{2}(r_i)}\right)^2
 f_B (r_i)dr_i}
\label{alpha}
\ee
where the integrals extend over the production region.
As we mentioned before, for concreteness we have fixed along 
this discussion  the neutrino magnetic moment $\mu_{\nu}=10^{-12} \mu_B$.

We will now obtain bounds on parameter $\alpha$ and the peak field $B_0$
for each profile derived from KamLAND data,
applying Gaussian probabilistic considerations to the global rate in the
whole energy range, $E_{\nu}=(2.6-8.125)~MeV$, and Poissonian considerations to the 
event content in the highest energy bins ($E_e> 6 $ MeV) where KamLAND 
observes zero events. We denote by $S^{0}_{\bar\nu}$ the event rate with
$B_{0}=10^7G$ for each given profile ($S^{0}_{\bar\nu}=S_{\bar\nu}(10^7G)$). 
Taking the number of observed events and subtracting the 
number of events expected from the best-fit oscillation 
solution [($\Delta m^2,\sin^2 2\theta)_{LMA}=(6.9\times 10^{-5}\ eV^2,1)$]
and interpreting this difference as a hypothetical 
signal coming from  solar antineutrinos, we have
\begin{eqnarray}
S^{sun}_{\overline{\nu}}&=& 
S_{obs}-S_{react}(LMA).
\end{eqnarray}
Inserting \cite{klstony} $S_{obs}=54.3\pm 7.5 $ and
$S_{react}(LMA)= 49\pm 1.3 $, we obtain
$ S_{obs}-S_{react}=\alpha
S^{0}_{\bar\nu} < 17.8\  (20.2)$ at 90 (95)\% CL. 
Within each specific profile it is seen from (\ref{alpha}) that the quantity 
$\alpha$ is simplified to $\alpha=(B_0/10^7G)^2$, so that the previous
inequality becomes  
\be
B_{0}^2<\frac{S^{sun}_{\overline{\nu}}}{S^{0}_{\bar\nu}}(10^7G)^2.
\ee
In this way we can derive for each given profile an upper bound on $B_0$.
The quantity $S^{0}_{\bar\nu}$ for profiles 1, 2 and 3 
and the respective upper bounds on $B_0$ are shown in table 1.  
These upper limits can be cast in a more general way if do not fix
the neutrino magnetic moment. To this end we will consider an arbitrary
reference value $\mu^{0}_{\nu}=10^{-12}\mu_{B}$. Then within each profile, $\alpha=
(\mu_{\nu}B_0/\mu^{0}_{\nu}~10^7G)^2$, where in the numerator and denominator
we have respectively the peak field value and some reference peak field 
value of the same profile. In the same manner as before we can derive the upper
bounds on $\mu_{\nu} B_0$ which are also shown in table 1.  

\vspace{0.6cm}

From the definition of $\alpha$ (\ref{alpha})  
it follows that the upper bounds on the antineutrino flux are independent of the 
field profile. These turn out to be $\phi_{\bar\nu}<0.0034\phi(^8B)$ and 
$\phi_{\bar\nu}<0.0038\phi(^8B)$ for 90 and 95\% CL respectively.

We can similarly and independently apply Poisson statistics to the five highest 
energy bins of the KamLAND experiment. No events are observed in this region and the
expected signal from oscillating neutrinos with LMA parameters is negligibly
small. We use the fact that the sum of Poisson variables of mean $\mu_i$ is
itself a Poisson variable of mean $\sum \mu_i$. The background (here the reactor
antineutrinos) and the signal (the solar antineutrinos) are assumed to be 
independent Poisson random variables with known means.
If no events are observed and in particular no background is observed, the
unified intervals \cite{cousins,Hagiwara:fs} $[0,\epsilon_{CL}]$ are $[0,2.44]$
at 90\% CL and $[0,3.09]$ at 95\% CL.

From here, we obtain $\alpha S_{\bar\nu}^{0} < \epsilon_{CL}$ or 
$\alpha < \epsilon_{CL}/S_{\bar\nu}^{0}$. Hence, as in the previous case,
we have
\be
B_{0}^2<\frac{\epsilon_{CL}}{S^{0}_{\bar\nu}}(10^7G)^2.
\ee
Using the expected number of events in the first 145 days of data taking 
and in this energy range $(6-8.125)~MeV$, we have derived upper bounds on 
$B_0$ (90 and 95\% CL) for all three profiles. They are shown in
table 2 along with the upper bounds on $\mu_{\nu} B_0$ taking $\mu_{\nu}$
as a free parameter.  
The antineutrino flux upper bounds are now $\phi_{\bar\nu}<0.0049\phi(^8B)$
$\phi_{\bar\nu}<0.0055\phi(^8B)$ at 90 and 95\% CL respectively. The KamLAND
expected signal for an arbitrary field profile corresponding to 95\% CL is
shown in fig. 3. 

The differences in magnitude among the bounds on $B_0$ and $\mu_{\nu}B_0$ 
presented in tables 1 and 2 for the different profiles 
are easy to understand. In fact, recalling that the $^8B$ production zone 
peaks at 5\% of the solar radius and becomes negligible at 
approximately 15\% (fig. \ref{fig1}), then
in order to generate a sizeable antineutrino flux, the magnetic 
field intensity should lie relatively close to its maximum in 
the range where the neutrino production is peaked. Thus for 
profile 1 the value of $B_0$ required to 
produce the same signal is considerably larger than for the other two,
while profile 3 is the most efficient one for antineutrino production.  


As referred to above, for different field profiles the probability curves will
differ only slightly in their shape if they lead to the same number of events.
In other words, for a given number of events the probability curves are essentially
the same, regardless of the field profile, a fact illustrated in fig. \ref{fig2}. As a
consequence, the energy spectrum of the expected solar antineutrino flux will
be nearly the same for any profile. In fig. \ref{fig4} we plot this profile independent 
spectrum together with the $^8 B$ one \cite{Bahc}, so that a comparison can be 
made showing the shift in the peak and the distortion introduced.

\begin{table}[h]
\begin{center}
\scalebox{0.70}{
\begin{tabular}{||c|c|c|c|c|c||} \hline \hline
     &     &      &     &     &    \\[-0.4cm]
 Profile & $S^{0}_{\bar\nu}(10^7G)$ & $B_{0}(90\% CL)$ & $B_{0}(95\% CL)$ & $\mu_{\nu}B_0(90\% CL)$ & $\mu_{\nu}B_0(95\% CL)$  \\
 & & $G$ & $G$ & $MeV$ & $MeV$  \\[0.1cm]  \hline
1.& $0.006$ & $5.27\times 10^8$ & $5.62\times 10^8$ & $3.05\times 10^{-18}$ & $3.25\times 10^{-18}$\\
2.& $0.137$ & $1.14\times 10^8$ & $1.21\times 10^8$ & $6.60\times 10^{-19}$ & $7.04\times 10^{-19}$\\
3.& $0.224$ & $8.92\times 10^7$ & $9.50\times 10^7$ & $5.16\times 10^{-19}$ & $5.50\times 10^{-19}$\\[0.1cm]
  \hline
\end{tabular} 
}
\end{center}
 \caption{\small  Solar antineutrino event rates, upper bounds on the peak field value
for $\mu_{\nu}=10^{-12}\mu_B$ and on $\mu_{\nu} B_0$ for arbitrary $\mu_{\nu}$ and $ B_0$,
assuming Gaussian statistics in the whole KamLAND spectrum.}
\end{table}

\vspace{-0.5cm}

\begin{table}[h]
\begin{center}
\scalebox{0.70}{
\begin{tabular}{||c|c|c|c|c|c||} \hline \hline
     &     &      &     &     &    \\[-0.4cm]
 Profile & $S^{0}_{\bar\nu}(10^7G)$ & $B_{0}(90\% CL)$ & $B_{0}(95\% CL)$ & $\mu_{\nu}B_0(90\% CL)$ & $\mu_{\nu}B_0(95\% CL)$  \\
 & & $G$ & $G$ & $MeV$ & $MeV$  \\[0.1cm]  \hline
1.& $0.004$ & $2.53\times 10^8$ & $2.85\times 10^8$ & $1.47\times 10^{-18}$ & $1.65\times 10^{-18}$\\
2.& $0.079$ & $5.56\times 10^7$ & $6.25\times 10^7$ & $3.22\times 10^{-19}$ & $3.62\times 10^{-19}$\\
3.& $0.130$ & $4.34\times 10^7$ & $4.88\times 10^7$ & $2.51\times 10^{-19}$ & $2.82\times 10^{-19}$\\[0.1cm]
   \hline
\end{tabular} 
}
\end{center}
\caption{\small  Same as table 1 assuming Poissonian statistics in the KamLAND
energy range $E_e=(6-8.125)~MeV$.}
\end{table}

\vspace{-5cm}

\begin{figure}[h]
\begin{center}
\hspace*{-1.6cm}
\epsfig{file=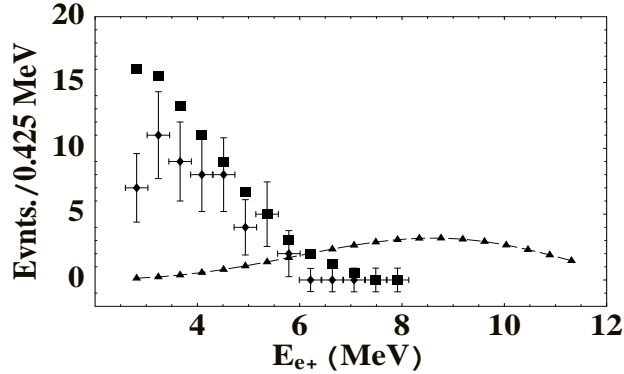,height=5.50cm,width=9.0cm,angle=0}
\end{center}
\caption{ \small 
The solid squares represent the MC expectation of 
the KamLAND positron spectrum from reactor antineutrinos with no 
oscillations and the points with error bars represent the measured
spectrum (from Fig.5 in Ref.\protect\cite{Eguchi:2002dm}).
Solid triangles represent the positron spectrum from solar
antineutrinos (multiplied by 5) assuming profile 3 with
peak field given by its 95\% CL upper limit ($B_0=4.88\times10^7G$).
All curves refer to the same time exposure of 145 days.} 
\label{fig3}
\end{figure}

\begin{figure}[h]
\begin{center}
\hspace*{-1.6cm}
\epsfig{file=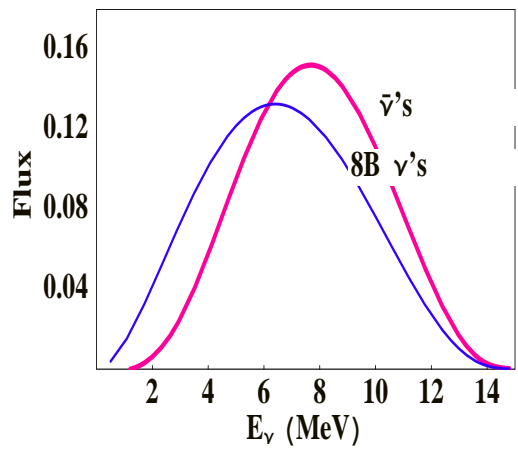,height=6.5cm,width=7.0cm,angle=0}
\end{center}
\caption{ \small The expected solar antineutrino spectrum and the 
$^8 B$ neutrino one \protect\cite{Bahc}, both normalized to unity, showing the 
peak shift and the distortion introduced by the antineutrino probability.}
\label{fig4}
\end{figure}

\clearpage

\section{  Conclusions}
\label{sec:conclusions}

In summary in this work we investigate  the possibility of 
detecting 
solar antineutrinos with the KamLAND experiment.
These antineutrinos are predicted by spin-flavor solutions
to the solar neutrino problem.

The KamLAND experiment is  sensitive 
to potential 
antineutrinos originated from solar ${}^8$B neutrinos.
We find that the 
results of the KamLAND experiment put  
relatively strict limits on the flux of solar electron antineutrinos
$\Phi( {}^8 B)< 1.1-3.5\times 10^4\ cm^{-2}\ s^{-1}$,
and their energy averaged appearance probability ($P<0.15-0.50\%$).
These limits are largely independent from any model on the solar 
magnetic field or any other astrophysical properties.
As we remarked in Section 2.1, these  upper 
limits on antineutrino probabilities and fluxes 
are still 
valid even if the antineutrino probabilities are significantly 
different from constant.

Next we assume a concrete model for antineutrino production
where they are produced  
by spin-flavor precession in the convective solar 
magnetic field. In this model, the antineutrino  
appearance probability is given by a simple expression as
$P(\overline{\nu})=\kappa\ \mu^2 \langle B^2\rangle$
with  $\kappa^{LMA}\simeq 2.8\times 10^{-44}$ MeV$^{-2}$.
In the context of this model and
assuming LMA central values for neutrino oscillation 
parameters
($\Delta m^2=6.9\times 10^{-5}$ eV$^2$, $\sin^2\theta=1$) 
\cite{kloctober}, the  upper 
 bound $\overline{p}< 0.15\%$ (95\% CL)
implies an upper limit on 
the product of the intrinsic neutrino magnetic moment and
the value of the convective solar magnetic field as 
$\mu\ B< 2.3\times  10^{-21}$ MeV (95\% CL).
For realistic values of other astrophysical solar
 parameters these upper 
limits would imply that the neutrino magnetic 
moment is constrained to be, in the most desfavourable case, 
$\mu\lsim 3.9\times 10^{-12}\ \mu_B$ (95\% CL) for a relatively small field $B= 50$ kG.  
For slightly higher values of the magnetic field:
 $\mu\lsim 9.0\times 10^{-13}\ \mu_B$ (95\% CL) for 
field $B= 200$ kG and 
 $\mu\lsim 2.0\times 10^{-13}\ \mu_B$ (95\% CL) for 
field $B= 1000$ kG. 
These assumed values for 
the  magnetic field at the base the solar convective zone 
are relatively mild and well within present astrophysical 
expectatives.

To conclude, now that SFP is ruled out as a dominant effect for the solar
neutrino deficit, it is important to investigate its still remaining possible
signature in the solar neutrino signal, namely an observable $\bar\nu_e$ flux.
Our main conclusion is that, from the antineutrino production model expound
here, an upper bound on the solar antineutrino flux can be derived, namely 
$\phi_{\bar\nu}<3.8\times 10^{-3}\phi(^8B)$ and $\phi_{\bar\nu}<5.5\times 
10^{-3}\phi(^8B)$ at 95\% CL, assuming respectively Gaussian or Poissonian 
statistics. For 90\% CL we found $\phi_{\bar\nu}<3.4\times 10^{-3}\phi(^8B)$
and $\phi_{\bar\nu}<4.9\times 10^{-3}\phi(^8B)$ which shows an improvement 
relative to previously existing bounds from LSD \cite{Aglietta} by a
factor of 3-5. These are independent of the detailed magnetic field profile 
in the core and radiative zone and the energy spectrum of this flux is also 
found to be profile independent. 
We also derive upper bounds on the peak field value which are 
uniquely determined for a fixed solar field profile. In the most
efficient antineutrino producing case (profile 3), we get  (95\% CL)
an upper limit on the product of the neutrino magnetic moment by the  
solar field $\mu_{\nu} B \leq 2.8\times 10^{-19}$ MeV or  
$B_0 \leq 4.9 \times 10^7 G$  for $\mu_\nu=10^{-12}\mu_B$.
A recent study of the magnetic field in the radiative zone of the sun 
has provided upper bounds of (3-7) MG \cite{Friedland:2002is} in that region in
the vicinity of $0.2~R_S$ which are independent
of any neutrino magnetic moment. Therefore we can use them in 
conjunction with our results to obtain a limit on $\mu_{\nu}$. 
Using $B_0\sim 3-7 MG$, we get from the results for profiles 1-3:  
$\mu~\lsim~0.7-9.6 \times 10^{-12}\mu_B$.
Moreover, from the limits obtained in this work, if the 'true' solar profile resembles 
either a profile like 1 or 3, this criterion implies that SFP cannot be experimentally 
traced in the next few years, since the peak field value must be substantially 
reduced in order to comply with this upper bound, thus leading to a much too 
small antineutrino probability to provide an observable event rate.
On the other hand, for a profile like 2 or in general any one resembling a dipole
field, SFP could possibly be visible.


\end{document}